\documentstyle[11pt]{article}
\def\rad{{\chi}}
\def\and{{$\&$ }}
\def\lsim{\mathrel{\mathpalette\oversim<}}
\def\gsim{\mathrel{\mathpalette\oversim>}}
\def\oversim#1#2{\lower0.2ex\vbox{\baselineskip0pt\lineskip0pt
  \lineskiplimit0pt\ialign{$#1\hfil##\hfil$\crcr#2\crcr\sim\crcr}}}
\begin{document}
\thispagestyle{empty}
{\baselineskip0pt
\leftline{\large\baselineskip16pt\sl\vbox to0pt{\hbox{Department of Physics}
               \hbox{Kyoto University}\vss}}
\rightline{\large\baselineskip16pt\rm\vbox to20pt{
           \hbox{KUNS-1357}
           \hbox{CfPA-95-TH-16}
           \hbox{August 1995}
\vss}}%
}
\vspace{2cm}
\begin{center}{\Large
  Observational Tests of One-Bubble Open Inflationary Cosmological Models}
\end{center}
\vspace{1.5cm}
\begin{center}
{\large Kazuhiro Yamamoto$~{}^{1}$ and ~Emory F. Bunn$~{}^{2}$} \\
\vspace{0.8cm}
{\em $~{}^{1}$~Department of Physics, Kyoto University} \\
{\em Kyoto 606-01,~Japan}\\
and\\
{\em $~{}^{2}$~Center for Particle Astrophysics, University of California,} \\
{\em Berkeley, CA 94720-7304, USA}\\
\end{center}
\vspace{1.0cm}

\begin{abstract}
Motivated by recent studies of the one-bubble inflationary
scenario, simple open cold dark matter models
are tested for consistency with
cosmological observations.
The initial perturbation spectrum is derived by solving for the
evolution of fluctuations in an open inflationary stage.
A likelihood analysis is performed for the Cosmic Microwave
Background anisotropies using the two-year COBE DMR data and considering
models based on both the Bunch-Davies and conformal vacua.
Having normalized the perturbation spectrum to fit the COBE data,
we reconsider the validity of the open
model from the view point of cosmic structure formation.
Open models may be severely constrained by the COBE likelihood
analysis.  In particular, small values of $\Omega_0$ are ruled
out in the Bunch-Davies case: we find that $\Omega_0\ge 0.34$
at 95\% confidence for this model.

\vspace{2mm}
\noindent
{{\it Subject headings}: cosmology: theory --- cosmic microwave background
--- large scale structure}
\end{abstract}
\vfill
\newpage
\section{Introduction}

The recent discovery of anisotropy in the cosmic microwave background (CMB)
lends support to the hypothesis that structure in the universe
formed via gravitational instability
from small density perturbations.  Furthermore, the observed anisotropy
allows us to characterize the initial perturbations statistically
with enough precision to draw important cosmological conclusions.
In particular, the COBE detection (Smoot et al.~1992, Bennett et al.~1994)
shows that
density perturbations existed on scales
larger than the horizon at the epoch of recombination.
Furthermore, the data are consistent
with the hypothesis that the perturbations are Gaussian distributed and
can be used to place constraints on the power spectrum of the initial
density perturbations.
CMB anisotropy is already one of the most important pieces of cosmological
data, and in coming years its importance will only increase.

In recent years, many cosmologists have favored theories in which
the density parameter $\Omega_0$ is equal to one.  In particular,
this value for $\Omega_0$ is predicted by
the simplest versions of the popular inflationary scenario.
However, many dynamical measurements indicate
a low-density universe (Peebles 1993, Ratra \and Peebles 1994a).
It is well known that the standard COBE-normalized Cold Dark Matter (CDM)
scenario, in which universe is flat and contains only baryons and
CDM particles, and in which the initial density perturbations
are of the Harrison-Zel'dovich type, predicts fluctuations that are
too large in amplitude on scales of galaxy clusters and below,
although a number of slight variants on this model can
be devised that fit the data better ({\it e.g.}, Bunn, Scott, \& White
1995; White et al.~1995).

There are two classes of low-density cosmological model
that can be motivated by the inflationary universe scenario.
One is the $\Lambda$-model. In this model
the matter density is small, so $\Omega_0<1$; however,
the cosmological constant contributes to the total mean
density required to make the universe flat: $\Omega_0+\Omega_\Lambda=1$.
The usual inflationary
scenarios predict that the universe is flat, so
that $\Omega_0+\Omega_\Lambda$ must be
extremely close to unity (Kashlinsky, Tkachev, \and Frieman 1994).
The cosmological constant is equal to the vacuum energy
density of the Universe.  Although we have no reason to be certain
that
it must be zero, the values that are of interest to cosmologists
({\it i.e.}, the values that make $\Omega_\Lambda$ of order unity)
are unnaturally small from the point of view of particle physics
(Weinberg 1989), and so these $\Lambda$-models are often regarded
as unappealing.

The other class of low-density inflationary scenario
is the open model, in which the universe has
negative spatial curvature. Recently, the possibility
of realizing an open universe has been discussed
in the context of inflation theory (Bucher, Goldhaber,
\and Turok 1994; Yamamoto, Sasaki, \and Tanaka 1995;
Linde 1995; Linde \and Mezhlumian 1995).
The essential idea is based on the semiclassical picture
of a bubble nucleation, which is described by a bounce
solution (Coleman \and De~Luccia 1980).
One bubble nucleation process can be regarded as the creation of
a homogeneous and isotropic spacetime with negative spatial
curvature inside the bubble due to the $O(4)$-symmetry of the
bounce solution.

It is of great interest to ask whether an open universe created
in this one-bubble inflationary scenario is observationally
acceptable or not. Open CDM models have been investigated
by many authors (Lyth \and Stewart 1990; Ratra \and Peebles
1994ab; Sugiyama \and Silk 1994; Kamionkowski et al. 1994;
G\'orski et al. 1995; Liddle et al. 1995).
Their investigations are based on the simple assumption
that the quantum state of a scalar field is in the conformal
vacuum state at the inflationary stage; however,
this is unlikely
to be the prediction of the one-bubble
inflationary scenario. It has been pointed out that
if we take the Bunch-Davies vacuum state as the state
of scalar field, the CMB
anisotropy in a low $\Omega_0$ universe appears quite different
due to the
super-curvature mode (Yamamoto, Sasaki, \and Tanaka 1995).

Following the usual inflationary picture in which
the quantum fluctuation of a scalar field generates
the density perturbation, we must investigate the
quantum state of fields inside the bubble.
Attempts have been made to study this problem
by developing a field-theoretical formalism based on
a multidimensional tunneling wave function
(Tanaka, Sasaki, \and Yamamoto 1994; Sasaki et al. 1994; Tanaka \and
Sasaki 1994; Yamamoto, Tanaka, \and Sasaki 1995; Hamazaki et al. 1995).
Bucher et al. have also considered this problem
(Bucher, Goldhaber \and Turok 1994; Bucher \and Turok 1995).
However, this problem requires further investigation.

In this paper, we consider the simple case in which the quantum
state of a scalar field is in the Bunch-Davies vacuum state,
and compare the predictions of this model with several cosmological
observations.
This situation is physically definite and clear, and as long
as
the bubble nucleation occurs in the de Sitter inflationary
background (where the initial inflationary period
has lasted sufficiently long), the field can be
approximated by the Bunch-Davies vacuum state, provided that
the effect of the bubble nucleation process is negligible.

In section 2, we first consider the initial spectrum by
solving for the evolution of cosmological perturbations in the
open inflationary stage.
In section 3 we use this initial power spectrum to
calculate various cosmological quantities and
compare these predictions
with observations.
We also compare these results with those of a previous
analysis of the open inflationary model based on the conformal vacuum state,
and a $\Lambda$-model with Harrison-Zel'dovich spectrum.
Section 4 is devoted to a discussion of our results.
We will work in units where $c=1$ and $\hbar=1$.

\section{Initial Conditions}
In this section, we consider the evolution of cosmological
perturbations in an open inflationary stage and derive the
initial spectrum of perturbations.
Ratra \and Peebles (1994b) have investigated the evolution of
cosmological perturbations in an open inflationary universe
with gauge fixed. Bucher, Goldhaber \and Turok (1994)
have also investigated cosmological perturbations in an
inflationary stage in the gauge-invariant formalism.
In the first half of this section, we follow the work by
Bucher, Goldhaber \and Turok (1994).

In an open universe the line element can be written as
\begin{equation}
  ds^2=a^2(\eta)\biggl[
  -d\eta^2+d\chi^2+\sinh^2\chi d\Omega^2_{(2)}\biggr],
\label{metric}
\end{equation}
where $\eta$ is conformal time. We consider a scalar
field $\phi$ with potential $V(\phi)$, which drives
inflation. Writing the scalar field as a homogeneous part and
a small inhomogeneous part, {i.e.}, $\phi=\phi_0+\delta\phi$,
the equations for the homogeneous part are
\begin{equation}
  \phi_0''+2{\cal H}\phi_0'+a^2 {\partial V\over\partial\phi}=0,
\end{equation}
and
\begin{equation}
  {\cal H}^2 -1={8\pi G\over 3}
  \biggl[{\phi_0'^2\over 2}+a^2V\biggr],
\end{equation}
where ${\cal H}:=a'/a$ and the prime
denotes differentiation with respect to $\eta$.

The small fluctuation $\delta\phi$ gives rise to a
metric perturbation. As we are interested in a
scalar perturbation in a scalar field dominated universe,
the metric perturbation can be written,
\begin{equation}
  ds^2=a^2(\eta)\biggl[
  -\bigl(1+2\Phi\bigr)d\eta^2+\bigl(1-2\Phi\bigr)
  \biggl(d\chi^2+\sinh^2\chi d\Omega^2_{(2)}\biggr)
\biggr].
\end{equation}
$\Phi$ corresponds  to the curvature perturbation or
gravitational potential. Then we get the evolution
equations for the perturbation
(Mukhanov, Feldman, \and Brandenberger 1992),
\begin{eqnarray}
&&\Phi''+2\Bigl({\cal H}-{\phi_0''\over\phi_0'}\Bigr)\Phi'
+\Bigl(-{\bf L}^2+4
+2{\cal H}'-2{\cal H}{\phi_0''\over\phi_0'}\Bigr)\Phi=0,
\label{eqPhi}
\\
&&\delta\phi={1\over4\pi G\phi_0'}
\bigl(\Phi'+{\cal H}\Phi\bigr),
\label{connect}
\end{eqnarray}
where
\begin{equation}
  {\bf L}^2:={1\over\sinh^2\chi}{\partial\over\partial\chi}
  \biggl(\sinh^2\chi{\partial\over\partial\chi}\biggr)+
  {1\over\sinh^2\chi}{\bf L}^2_\Omega,
\end{equation}
and ${\bf L}^2_\Omega$ is the Laplacian on the unit sphere.

Now let us consider an inflationary stage of the universe
inside a bubble.  To solve the above equations analytically,
we assume that the potential is nearly flat, and
use the approximation, $V\simeq V_0+V'\phi$,
where $V'={\rm const}$.
We also assume that the background spacetime is approximated
by de Sitter spacetime, that is, $a(t)=-1/H\sinh\eta$.

Then the field equation for the homogeneous part is
\begin{equation}
  \phi_0''-2{\rm coth}\,\eta\phi_0' =
  {-V' \over H^2\sinh^2\eta},
\label{phi0eq}
\end{equation}
with $H^2={8\pi G V_0/3}$.

Eq. (\ref{phi0eq}) can be integrated, giving
\begin{equation}
\phi_0'={-V'\over H^2}
{-\cosh^3\eta+3\cosh\eta\sinh^2\eta+2\sinh^3\eta\over 3\sinh\eta},
\end{equation}
and the perturbation equation (\ref{eqPhi}) reduces to,
\begin{equation}
\Phi''-{6(1-e^{2\eta})\over 3-e^{2\eta}}\Phi'+
\biggl(-{\bf L}^2 +4-{4(3+e^{2\eta})\over3-e^{2\eta}}\biggr)
\Phi=0.
\label{eqPhib}
\end{equation}

To solve these equations, we need the initial values.
To determine the initial values, we must investigate
the problem of what the quantum state is inside the bubble.
As mentioned before, this is a very important problem,
which demands further investigation.
We here consider the case in which the scalar field
is in the Bunch-Davies vacuum state.
This is the case provided that the effect of bubble
nucleation process is small and negligible. But we should
keep in mind that this point needs examination in the various
models of one-bubble inflation scenario, taking into consideration
the effect of
bubble nucleation.

Recently, quantum field theory in de Sitter space-time
associated with the open chart has been investigated
(Sasaki, Tanaka, \and Yamamoto 1995). According to this analysis,
a quantized scalar field with mass $m^2\ll H^2$ in the
Bunch-Davies Vacuum state is described in the second
quantized manner as,
\begin{equation}
  \delta\phi=\int_0^\infty dp\sum_{\sigma,l,m}
  \chi_{p,\sigma}(\eta)Y_{plm}(\rad,\Omega_{(2)})\hat a_{p\sigma lm} +
  \sum_{lm}v_{(*)lm}(t,\rad,\Omega_{(2)}) \hat a_{(*)lm} +
  {\rm h.c.},
\end{equation}
where
\begin{eqnarray}
  \chi_{\sigma,p}(\eta)&=&{-1\over\sqrt{8p(p^2+1)\sinh\pi p}}
\nonumber
\\
  &&\biggl[e^{\pi p/2}(ip+\coth\eta)e^{-ip\eta}
  +\sigma e^{-\pi p/2}(ip-\coth\eta)e^{ip\eta}
  \biggr]{1\over a(\eta)},
\label{chi}
\end{eqnarray}
\begin{eqnarray}
  v_{(*)lm}(t,\chi,\Omega_{(2)})
&=&{H\over2}\sqrt{\Gamma(l+2)\Gamma(l)}
  {P^{-l-1/2}_{1/2}(\cosh\chi)\over \sqrt{\sinh\chi}}
  Y_{lm}(\Omega_{(2)})
\nonumber
\\
&=:&{H\over2}W_{(*)l}(\chi)Y_{lm}(\Omega_{(2)}),
 {\hspace{14mm}} (l>0),
\label{vstar}
\end{eqnarray}
$\sigma$ takes on the values $\pm1$, $\hat a$ is the annihilation
operator, and $\Gamma(z)$ is the gamma function.
The orthonormal harmonics on a three-dimensional unit
hyperboloid $Y_{plm}(\chi,\Omega)$ are
\begin{equation}
  Y_{plm}(\chi,\Omega_{(2)})
  =\Bigg\vert{p\Gamma(ip+l+1)\over\Gamma(ip+1)}\Bigg\vert
  {P^{-l-1/2}_{ip-1/2}(\cosh\chi)\over \sqrt{\sinh\chi}}
  Y_{lm}(\Omega_{(2)}),
\end{equation}
with normalization
\begin{equation}
  \int_0^\infty d\chi\int d\Omega_{(2)}
  \sinh^2\chi Y_{p_{1}l_{1}m_{1}}(\chi,\Omega_{(2)})
  \overline{Y_{p_{2}l_{2}m_{2}}(\chi,\Omega_{(2)})}
  =\delta(p_{1}-p_{2}) \delta_{l_{1}l_{2}} \delta_{m_{1}m_{2}}.
\end{equation}
In the above expression, the usual harmonics behave as
$Y_{plm}\propto e^{-\chi}$ at scales larger than the
curvature scale, $\chi\gg1$, although $v_{(*)lm}$ is constant for
$\chi\gg1$. Thus $v_{(*)lm}$ represents a fluctuation
larger than the curvature scale, so we call this mode a
super-curvature mode. The necessity of super-curvature modes
for a complete description of a random field in an open
universe has also been discussed by Lyth \and Woszczyna
(1995).

Next, let us consider the curvature perturbation,
which can be written in the mode-expanded form
\begin{equation}
\Phi=\int_0^\infty dp \sum_{\sigma,l,m}
\Phi_{p,\sigma}(\eta) Y_{plm}(\chi,\Omega_{(2)})+
\sum_{l,m}\Phi_{(*)}(\eta)W_{(*)l}(\chi)Y_{lm}(\Omega_{(2)}).
\end{equation}

For the continuous mode $(p,l,m)$, Eq.(\ref{eqPhi})
reduces to
\begin{equation}
\Phi_p''-{6(1-e^{2\eta})\over 3-e^{2\eta}}\Phi_p'+
\biggl(p^2 +5-{4(3+e^{2\eta})\over3-e^{2\eta}}\biggr)
{\Phi_p}=0.
\end{equation}
The solution that behaves like
$\tilde\Phi_p\rightarrow e^{(1-ip)\eta}$ as
$\eta\rightarrow -\infty$ is (Bucher \and Turok 1995)
\begin{equation}
 \tilde\Phi_p=e^{(1-ip)\eta}\biggl(1+{1+ip\over 1-ip}
{e^{2\eta}\over3}\biggr).
\end{equation}
The equation for the super-curvature mode is
\begin{equation}
\Phi_{(*)}''-{6(1-e^{2\eta})\over 3-e^{2\eta}}\Phi_{(*)}'+
\biggl(4-{4(3+e^{2\eta})\over3-e^{2\eta}}\biggr)
{\Phi_{(*)}}=0,
\end{equation}
and we find a solution
\begin{equation}
 \tilde\Phi_{(*)}=e^{2\eta}.
\end{equation}

To determine the amplitude of $\Phi$, we use
Eq.(\ref{connect}). From the behavior at
$\eta\rightarrow-\infty$, we find for the continuous mode
\begin{equation}
  \Phi_{p,\sigma}(\eta)={2\pi G V'\over H}
  {1\over\sqrt{8p(p^2+1)\sinh\pi p}}
  \biggl\{
  {e^{\pi p/2}}{1-ip\over2-ip}
  \tilde\Phi_{p}
  +\sigma
  {e^{-\pi p/2}}{1+ip\over2+ip}
  \tilde\Phi_{-p}\biggr\},
\end{equation}
and for the super-curvature mode
\begin{equation}
  \Phi_{(*)}(\eta)={2\pi G V'\over H}
  {1\over3}\tilde\Phi_{(*)}(\eta).
\end{equation}

We therefore have the following spectrum at the
end of inflation, by taking the limit $\eta\rightarrow0$,
\begin{eqnarray}
  \Phi_p(0)^2
  &:=&\lim_{\eta\rightarrow0}
  \sum_{\sigma=\pm1} \Phi_{p,\sigma}(\eta)^2
\nonumber
\\
  &=&\biggl({2\pi G V'\over H}\biggr)^2{\coth\pi p\over 2p(p^2+1)}
  {p^2+1\over p^2+4}
  \lim_{\eta\rightarrow0}
  \Bigl\vert \tilde\Phi_p(\eta)\Bigr\vert{}^2
\nonumber
\\
  &=&\biggl({4\pi G V'\over 3H}\biggr)^2
  {\coth\pi p\over 2p(p^2+1)},
\label{specnm}
\end{eqnarray}
and
\begin{eqnarray}
  \Phi_{(*)}(0)^2
  &=&\lim_{\eta\rightarrow0}\Phi_{(*)}(\eta)^2
\nonumber
\\
  &=&\biggl({4\pi G V'\over 3H}\biggr)^2{1\over4}.
\label{specsc}
\end{eqnarray}

For comparison, we also investigate the case when
the scalar field is assumed to be in the conformal
vacuum state (Lyth \and Stewart 1990, Ratra \and Peebles
1994b). In this case, the scalar field is written
\begin{equation}
  \delta\phi=\int_0^\infty dp\sum_{l,m}
  \chi_{p}(\eta)Y_{plm}(\rad,\Omega_{(2)})\hat b_{plm}
  +{\rm h.c.} ~,
\end{equation}
where
\begin{equation}
  \chi_{p}(\eta)={(ip+\coth\eta)e^{-ip\eta}\over\sqrt{2p(p^2+1)}}
{1\over a(\eta)}.
\end{equation}
After a similar analysis, we get the
spectrum of curvature perturbations at the end of inflation,
\begin{equation}
  \lim_{\eta\rightarrow0}\Phi_p(\eta)^2
  =\biggl({4\pi G V'\over 3H}\biggr)^2  {1 \over 2p(p^2+1)}.
\label{spev}
\end{equation}
The conformal vacuum case differs from the
Bunch-Davies vacuum case in two ways, the factor $\coth\pi p$
and the super-curvature mode.

Lyth \and Stewart (1990) have investigated perturbations
in an open inflationary universe, and have given a relation
to relate the curvature perturbation and the scalar field
perturbation ${\cal R}\simeq -(H/\dot\phi)\delta\phi$,
though a paper justifying this relation has never been published.
But the above investigation shows the correctness of their result
on all scales except for the small difference of the
former coefficient.

\section{Observational Confrontations}
Now we start testing the predictions of the open universe
in the context of CDM cosmology with the initial conditions
obtained above. The matter-dominated open universe
has the line element (\ref{metric}) with $a(\eta)=\cosh\eta-1$.
In this section we use $\eta(>0)$ as the conformal time in the
matter-dominated universe.

\begin{center}
\underline{(1) CMB Anisotropies}
\end{center}
Let us first consider the CMB temperature fluctuation.
Having obtained an initial perturbation spectrum, we can
compute the temperature fluctuations in the gauge-invariant
formalism (Sugiyama \and Gouda 1992).
As usual, we write the temperature
autocorrelation in the form,
\begin{equation}
  C(\alpha)={1\over4\pi}\sum_{l}(2l+1)C_l{\rm P}_l(\cos\alpha).
\end{equation}
Figure 1(a) shows the power spectrum of temperature fluctuations,
$l(l+1)C_l\times10^{10}/2\pi$, for various values of $\Omega_0$
with initial conditions associated with the Bunch-Davies
vacuum state. We have taken $\Omega_B h^2=0.0125$ and Hubble
parameter $h=0.75$, $0.70$, $0.65$, $0.65$, $0.60$, for
$\Omega_0=0.1$, $0.2$, $0.3$, $0.4$, $0.5$, respectively,
to take the age problem into consideration.
Figure 1(b) shows the same quantities with the
conformal vacuum state (Kamionkowski et al. 1994; G\'orski et al. 1995).
For reference, we show the corresponding results for a $\Lambda$-model with
a Harrison Zel'dovich spectrum in Figure 1(c) (Sugiyama 1995).
Here the parameter $\Omega_B h^2=0.0128$ and $h=0.8$.
We also show the results of several CMB experiments, taken
from the paper by Scott, Silk \and White (1995).
Open models may have trouble fitting the data near the ``Doppler
peak'' on degree scales, although assessing the significance of this
problem will require very careful investigation
(Ratra et al. 1995).

The differences between Fig.1(a) and Fig.1(b) at low multipoles
come almost entirely from the contribution of the super-curvature mode
(Yamamoto, Sasaki, \and Tanaka 1995),
\begin{equation}
  C_{(*)l}=\biggl({2\pi G V'\over 3H}\biggr)^2
  \biggl\{{1\over3}f(\eta_{LS})W_{(*)l}(\eta_0-\eta_{LS})
  +2\int_{\eta_{LS}}^{\eta_0}d\eta' {df(\eta')\over d\eta' }
  W_{(*)l}(\eta_0-\eta') \biggr\}^2,
\end{equation}
where $\eta_{LS}$ and $\eta_0$ are the recombination time and the present
time, respectively, $W_{(*)l}$ is defined in Eq.(\ref{vstar}), and
$f(\eta)$ is the decay factor of the curvature perturbation,
\begin{equation}
  f(\eta)=5~{\sinh^2\eta-3\eta\sinh\eta+4\cosh\eta-4\over(\cosh\eta-1)^3}.
\label{feta}
\end{equation}

The most accurate and reliable CMB anisotropy data at the present
time come from the COBE DMR experiment.  In addition to providing us
with accurate estimates of the fluctuation amplitude, the data from
this experiment can be used to constrain the shape of the power spectrum.
We have used the two-year COBE data (Bennett et al.~1994)
to place constraints on open inflationary
models, following a procedure based on the Karhunen-Lo\`eve transform
(Bunn, Scott, \& White 1994; Bunn \& Sugiyama 1995; White \& Bunn 1995;
Bunn 1995ab).  This procedure gives results that are generally
consistent with the spherical-harmonic technique devised by G\'orski
(1994).
We will now describe this procedure.

In inflationary cosmological models, the CMB anisotropy is a realization
of a Gaussian random field.  If we expand the anisotropy in spherical
harmonics,
\begin{equation}
\Delta T(\hat{\bf r}) = \sum_{l=2}^\infty\sum_{m=-l}^l a_{lm}Y_{lm}
(\hat{\bf r}),
\end{equation}
then each coefficient $a_{lm}$ is an independent Gaussian random variable
of zero mean.  Furthermore, the variance of $a_{lm}$ is simply $C_l$.
With this information, we can in principle compute the probability
density $p(\vec d \,|\, C_l)$ of getting the actual COBE data $\vec d$ given
a power spectrum $C_l$: since each data point is a linear combination
of Gaussian random variables, the probability distribution $\vec d$
is simply a multivariate Gaussian,
\begin{equation}
p(\vec d \,|\, C_l)\propto\exp\left(-{1\over 2}\vec d^TM^{-1}\vec d\right),
\label{GaussProb}
\end{equation}
where the covariance matrix $M$ can be written in terms of the
power spectrum and the noise covariance matrix.

Let us restrict our attention to a few-parameter family of possible
power spectra.  We will denote the parameters generically by $\vec q$.
In this paper, for example, we will consider
power spectra that are parameterized by two parameters, the
density parameter $\Omega_0$ and the power spectrum normalization
$Q\equiv\sqrt{5C_2/4\pi}$, and so $\vec q$ will be a two-dimensional
vector.  The probability density in
equation (\ref{GaussProb}) is then simply the probability density $p(\vec d\,|
\,\vec q)$
of the data $\vec d$ given the parameters $\vec q$.
If we adopt a Bayesian view of statistics,
we can convert this into a probability density for the parameters
given the data:
\begin{equation}
p(\vec q \,|\, \vec d)\propto p(\vec d \,|\, \vec q) p(\vec q),
\end{equation}
where $p(\vec q)$ is the prior probability density we choose to
adopt.  $p(\vec q \,|\,\vec d)$ is generally denoted $L(\vec q)$
and called the likelihood.

The choice of prior distribution is a notoriously troublesome issue.
In practice, one generally chooses a prior
that is a smooth, slowly-varying function of the
parameters.  In this paper, we will adopt a prior that is uniform
in $\Omega_0$ and one that is uniform in $\ln Q$.  (This prior
is approximately equivalent to one that is uniform in the power spectrum
normalization $C_{10}$ near the ``pivot point.''  It differs slightly
from one that is uniform in $Q$, although not enough to affect our
results significantly.)

Unfortunately, in order to compute the probability density $p(\vec
d\,|\,\vec q)$, and hence the likelihood $L$, one must invert a matrix of
dimension equal to the number of data points.  For the COBE DMR data,
this number is of order 4000.  Such exact likelihoods have been
computed for a small class of models (Tegmark \& Bunn 1995); however,
this is quite a time-consuming procedure.  The Karhunen-Lo\`eve
transform allows us to ``compress'' the data from 4000 numbers to only
400 in a way that throws away very little of the actual cosmological
signal.  The likelihoods estimated from the transformed data
approximate the true likelihoods well, and are much more efficient to
compute.  For details on how the Karhunen-Lo\`eve transform is
performed, see White \& Bunn (1995) and Bunn (1995ab).

Once we know $L$, it is quite easy to place constraints on the parameters
$\vec q$.  Since $L$ is a probability distribution for $\vec q$, it
should be normalized so that
\begin{equation}
\int L(\vec q)\,d\vec q = 1.
\end{equation}
Now suppose that we choose some subset $R$ of possible parameter values.
Then if
\begin{equation}
\int_R L(\vec q)\,d\vec q=c
\end{equation}
then we can say that $\vec q$ lies in the region $R$ with probability
$c$.  If we want to find a 95\% confidence interval, we simply
find a region $R$ such that $c=0.95$.  One frequently chooses $R$
to be the region enclosed by a contour of constant likelihood.

If one of the parameters is deemed to be uninteresting, the standard
practice is to ``marginalize'' over it.  For example, if we are interested
in constraining $\Omega_0$ but not $Q$, then we replace $L(\Omega_0,Q)$
by
\begin{equation}
L_{\rm marg}(\Omega_0)=\int L(\Omega_0,Q)\,dQ.
\end{equation}
This is a natural thing to do: if $L$ is the joint probability density
for $\Omega_0$ and $Q$, then $L_{\rm marg}$ is the probability density
for $\Omega_0$ alone.

Figure 2 shows the contours of the likelihood $L(\Omega_0,Q)$ for open models
associated with the Bunch-Davies vacuum state.  In computing these
likelihoods, we use a linear combination of the 53 and 90 GHz
two-year COBE maps, with weights chosen to minimize the noise.  We
use the ecliptic-projected maps; maps that were made in Galactic
coordinates, and therefore have different pixelization, give normalizations
that are generally lower by a few percent (Stompor, G\'orski \& Banday
1995; Bunn 1995a).
The choice of pixelization
appears to affect primarily the overall normalization of models;
likelihood ratios of models with power spectra of different shapes
are less affected (Bunn 1995ab).

Figure 3 shows the marginal likelihoods for $\Omega_0$ for both
the Bunch-Davies and the conformal vacuum open models.
In the Bunch-Davies case, we find that $\Omega_0>0.34$ at 95\%
confidence and $\Omega_0>0.15$ at 99\% confidence.
We also show the confidence levels for various $\Omega_0$
in Table 1.
For the conformal vacuum models, the likelihood is bimodal,
and so the allowed regions are not connected.
If we take a cut at small $\Omega_0$ and only consider the region
$\Omega_0\geq0.03$, we can state that at 95\% confidence
either $\Omega_0<0.085$ or $\Omega_0>0.36$, and  99\% confidence,
either $\Omega_0<0.14$ or $\Omega_0>0.23$.
There are difficulties associated with the interpretation of the
likelihoods in this case
(G\'orski et al. 1995).

Of course, the likelihood $L(\Omega_0,Q)$ provides us with accurate
normalizations in addition to shape constraints.  For any particular
value of $\Omega_0$, we find the value of $Q$ that maximizes the
likelihood and use this value as the power spectrum normalization.
The normalizations determined in this way have typical one-sigma
fractional uncertainties of approximately $7.5\%$.
The maximum-likelihood normalizations computed in this way
are listed in the second column of Table 2(a) for the
Bunch-Davies vacuum case. For comparison, we have also
computed the conformal vacuum case; these normalizations are given in
the
second column of Table 2(b).


\vspace{3mm}
\begin{center}
\underline{(2) Linear density power spectrum}
\end{center}
We next consider the matter inhomogeneities
using the COBE normalization as described above. As the density
perturbation $\Delta$ is related to the curvature perturbation
$\Phi$ by the gravitational Poisson equation (Kodama \and Sasaki 1984),
\begin{equation}
  (p^2+4)\Phi_p(\eta)=4\pi G\rho(\eta) a^2(\eta)\Delta_p(\eta),
\end{equation}
in linear perturbation theory, we can write
the power spectrum of the matter perturbation in an open
universe from Eq.(\ref{specnm}),
\begin{eqnarray}
  a_0^3P(k)~\Bigl(:=a_0^3\Delta_p^2\Bigr)
&=&\biggl({2(1-\Omega_0)\over 3\Omega_0}\biggr)^2
  (p^2+4)^2a_0^3\Phi_p^2(0) f^2(\eta_0) T(k)^2
\nonumber
\\
&=:&{\cal A}(p^2+4)^2{\coth \pi p \over p(p^2+1)}T(k)^2,
\label{Pk}
\end{eqnarray}
where  $p=a_0 k$, $a_0=1/H_0\sqrt{1-\Omega_0}$,
and $H_0=100h{\rm km/s/Mpc}$.
The super-curvature mode does not contribute
on small scales. The model based on the
conformal vacuum state leads to the same form
but without the factor $\coth\pi p$.
In the CDM cosmology, the following transfer function is useful
(Bardeen et al. 1986, Sugiyama 1995),
\begin{equation}
  T(k)={\log(1+2.34q)\over 2.34q}
  \Bigl[1+3.89q+(16.1q)^2+(5.46q)^3+(6.71q)^4\Bigr]^{-1/4},
\label{transfersu}
\end{equation}
with
\begin{equation}
  q=\Big({2.726\over2.7}\Bigr)^2
  {k\over\Omega_0h\exp(-\Omega_B-\sqrt{2h}\Omega_B/\Omega_0)}
  ~h{\rm Mpc}^{-1} .
\end{equation}
This transfer function is for a flat universe; however, since we are
interested in small-scale perturbations, the
curvature of the universe can be neglected and the
transfer function above is acceptable.

The COBE DMR normalization determines the amplitude of the
fluctuation. We give the numerical value of ${\cal A}$
determined from the likelihood normalization
in the second column in Table 2(a).
The second column in Table 2(b) gives the value for
the conformal vacuum case, in which the power spectrum
is obtained by Eq.(\ref{Pk}) without the factor $\coth\pi p$.

We show in Figure 4(a) the density power spectrum $a_0^3P(k)$,
for various $\Omega_0$, with initial conditions
based on the Bunch-Davies vacuum state. The Hubble parameter
is the same as that in Fig.1(a).
The points are from Peacock \and Dodds (1994).
Figure 4(b) shows the density power spectra for $\Lambda$-models with
Harrison-Zel'dovich power spectrum, in which the
Hubble parameter is same as that in Fig.1(c).

Given the density perturbation spectrum $P(k)$, we are able to
calculate $\sigma_8^2=$$(\delta M$$/M)^2_{8 h^{-1}{\rm Mpc}}$,
the variance of the mass fluctuation in a
sphere of a radius $R=8h^{-1}{\rm Mpc}$,
\begin{equation}
  \sigma^2(R)={1\over2\pi^2}\int k^2dk P(k) W^2(kR),
\end{equation}
where the top-hat window function $W$ is defined by
$W(x)=3(\sin x -x\cos x)/x^3$.
In Tables 2(a) and 2(b), we give
$\sigma_8$ for various $\Omega_0$ and $ h$ in the open model
associated with Bunch-Davies vacuum and the conformal vacuum respectively.
We also show values for the $\Lambda$ model
in Table 2(c) (Sugiyama 1995; Stompor et al. 1995).

The difference between the Bunch-Davies case
and the conformal vacuum case is very small. The
difference is $10$ percent at $\Omega_0=0.05$, but
is a few percent even at $\Omega_0=0.1$. This is because
the COBE normalization based on a likelihood analysis gives
more weight to the behavior of the power spectrum at $l\simeq10$,
and less weight at lower multipoles (White \and Bunn 1995).
When we take a $\sigma(10^\circ)$ normalization,
there is a $10$ percent difference between the two cases
at $\Omega_0=0.1$.
The values obtained are consistent with those in G\'orski et al.
(1995).

A precise comparison between predictions and observations
of the matter power spectrum is difficult.  One of the primary
problems is that we do not know whether the galaxy distribution is
an unbiased tracer of the mass distribution.  However, if we
make the reasonable assumptions that the galaxies are not anti-biased
and are not extremely strongly biased (say $b\equiv\sigma_8^{-1}\lsim
2.5$), then
these calculations suggest that $0.3\lsim\Omega_0\lsim0.5$. Note
that the values of $\Omega_0$ preferred by the COBE likelihood
analysis tend to be higher than this range; however, one might
be inclined to argue that a model with $\Omega_0\simeq 0.4-0.5$
passes both tests.

\vspace{3mm}
\begin{center}
\underline{(3)Large-scale bulk velocity}
\end{center}
Next, we consider large-scale bulk velocities,
which are given by the following expression,
\begin{equation}
  v_R^2= {H_0^2a_0^2\over 2\pi^2} \Omega_0^{1.2} \int dk P(k) W(kR)^2
  \exp(-k^2R_s^2),
\end{equation}
where $W(kR)$ is the window function, and $R_s=12h^{-1}{\rm Mpc}$
is the Gaussian smoothing length for comparison with the observational
data.
In Table 3, we have summarized the computation of $v_R^2$
with $R=40h^{-1}{\rm Mpc}$ for various $\Omega_0$ for open
models with the initial conditions associated with the Bunch-Davies vacuum
state. These results are consistent with those of G\'orski et al. (1995).

We can compare this results with the recent data from the POTENT
analysis (Dekel~1994; Liddle et al.~1995):
$v_{R=40h^{-1}{\rm Mpc}}=373\pm50{\rm km/s}$.
Large values of $\Omega_0$ clearly provide a better fit to the velocities.
It appears difficult
to reconcile models with $\Omega_0\lsim 0.3$ with these data; however,
it is difficult to make precise statistical statements based on these
observations.
As is discussed by Liddle et al. (1995),
this measurement of the bulk velocity contains additional uncertainty due to
cosmic variance.  In addition, it is quite difficult to assess the
uncertainties and potential biases in the POTENT analysis, and
one should therefore be reluctant to draw firm conclusions
on the basis of such a comparison.

\vspace{3mm}
\begin{center}
\underline{(4) epoch of galaxy formation}
\end{center}
Liddle et al. (1995) have performed a detailed
investigation of abundances of galaxy clusters
and damped Lyman-alpha systems in open CDM models using
Press-Schechter theory.
In this paper we will rely on
a simple and rough estimate of the epoch of galaxy formation,
following the work of Gottl\"ober, M\"uchet, $\&$ Starobinsky (1994)
and Peter, Polarski, $\&$ Starobinsky (1994).
According to Press-Schechter theory, the fraction of the matter in
the universe which is in gravitationally bound objects above
a given mass $M_R$ at a redshift $z$ has the form
\begin{equation}
  F(>M_R)={\rm erfc}\Bigl({\delta_c\over\sqrt{2}\sigma(M_R,z)}\Bigr),
\end{equation}
where
\begin{equation}
  \sigma(M_R,z)=\sigma(R){1\over 1+z}
  {f(\eta_0)\over f(\eta(z))},
\end{equation}
where $M_R=(4/3)\pi R^3\rho$, and $f(\eta)$ is the decay
factor of the curvature perturbation.

The choice of $\delta_c$ depends on the collapse model.
The spherical collapse of a top-hat perturbation
gives $\delta_c=1.69$, although non-spherical collapse models
suggest other values. Here let us consider the range
$(1.33<\delta_c<2)$ (Gottl\"ober et al. 1994).
Observations suggest that many galaxies seems to have formed at
$z=1$, then, assuming $F(>10^{12}{\rm M}_\odot) \gsim 0.1$ at $z=1$, we have
$\sigma(M_R=10^{12}{\rm M}_\odot,z=1)\gsim 2\pm0.4$.

Figure 5(a) shows a contour plot of $\sigma(M_R=10^{12}{\rm M}_\odot,z=1)$
in the $\Omega_0-h$ plane for the open model. Figure 5(b) is same but
for the $\Lambda$-model with a Harrison Zel'dovich spectrum.
If we take the age problem into consideration, it indicates
a lower bound $\Omega_0\gsim 0.4$ for the open model.
Note that the above estimate is very rough,
although a similar constraint has been obtained from
the exact estimation of cluster abundances (Liddle et al. 1995).
The bound is weaker in the $\Lambda$-model than in the open model.

\section{Discussion}
A low-density universe is well motivated from several dynamical
observations of galaxies and clusters. The simplest such
low-density models are those in which the universe is open.
In the context of inflation theory, however, we need a special idea
such as the
one-bubble inflationary scenario in order to produce an open universe.
In this paper, motivated by the one-bubble inflationary universe
scenario, we have examined the cosmological predictions based on the
assumption that the scalar field is initially in
the Bunch-Davies vacuum state.
The initial perturbation spectrum has been derived by
considering the evolution of perturbations in an open inflationary
stage. Then the CMB anisotropies and the matter inhomogeneities
have been examined.

As the first test, we have performed a likelihood analysis
for the CMB anisotropies by using the COBE DMR data.
Interestingly, the COBE likelihood analysis
gives severe constraints on the model.  Models with
$\Omega_0\leq0.4$, $\Omega_0\leq0.5$ are excluded at confidence
levels of
$92\%$, $83\%$, respectively.
In a previous analysis associated with the conformal
vacuum state (G\'orski et al. 1995), the likelihood function
has another steep peak below $\Omega_0\lsim0.15$.
This complicates the statistical interpretation of the results
(G\'orski et al. 1995).
In the case of the Bunch-Davies vacuum state, no such peak
appears in the range of $\Omega_0$ we are interested in, and so the
likelihood analysis gives clear results.
The COBE likelihood analysis is therefore a powerful probe of
these open models.

We have used the
the COBE DMR maximum-likelihood normalization to predict the
amplitude of matter fluctuations.  According to this
normalization method, there is little difference between the predictions
of the Bunch-Davies vacuum and conformal vacuum cases.  Even for
the case $\Omega_0=0.1$, the discrepancy of $\sigma_8$
is a few percent.  We obtain results that are similar to previous
analyses:
the power spectrum of the mass fluctuation fits the observations of
galaxies and clusters for
$0.3\lsim\Omega_0\lsim0.5$.
The required bias is unacceptably high
for $\Omega_0\lsim0.1$, while high values of $\Omega_0$
demand anti-biasing. For example,  $\Omega_0\gsim0.6$
needs anti-biasing when $h=0.65$.
On the other hand, the $\Lambda$-models with Harrison Zel'dovich
spectrum have higher amplitude compared with open
models. The $\Lambda$-model needs anti-biasing for $\Omega_0\gsim0.4$
even when $h=0.65$.
The $\Lambda$-model therefore needs low $h$ or a tilted spectrum
(Ostriker \and Steinhardt 1995).
For the range $0.3\lsim\Omega_0\lsim0.5$, open models
give unacceptably small bulk velocities compared with the
POTENT analysis. However, given the present quality of the velocity
data and
the problem of cosmic variance, one might be reluctant to draw
strong conclusions from this fact.
The rough estimation associated with the
galaxy formation gives lower bound of $\Omega_0$
consistent with the value discussed above.

It is very interesting that the COBE likelihood analysis
has given the most severe constraint on this open model.
The COBE likelihood analysis strongly prefers a high value of $\Omega_0$.
The peak value is around $0.7\lsim\Omega_0\lsim0.8$, and we can
state that
$\Omega_0\geq0.5$ with $83\%$ confidence in this model.
Considering both the COBE analysis and the matter inhomogeneity,
we are led to prefer a value of $\Omega_0\simeq 0.5$ if the
one-bubble inflationary scenario is correct.  Such a model is
consistent with the Press-Schechter analysis of the epoch of galaxy
formation and is marginally consistent with the bulk velocity data.
As the CMB data continue to improve, particularly on degree scales,
we should be able to test this model.

It is premature to rule out low $\Omega_0$ inflationary models on the
basis of this
investigation at present, because we do not include
the effect of bubble nucleation in the calculation of
initial density power spectrum.
Previous analysis indicates that the bubble nucleation effect
in general excites fluctuations, and amplifies the perturbations on
scales larger than curvature scale (Yamamoto, Tanaka, \and Sasaki 1995;
Hamazaki et al. 1995). One might therefore expect low-density models
to fit the COBE data even more poorly once this effect is taken
into account; however, since the calculation has not been done, we
cannot be certain.  In particular, the status of the super-curvature
mode is still quite uncertain.

Various modifications of the open model may also be viable.
We must investigate the effect of gravity waves
in an open inflationary universe.  One might also consider the effect of
tilting the primordial power spectrum; however, in order
to improve the fit to the data one would probably
need to tilt the power spectrum to have increased power on small scales,
and such ``blue'' power spectra are not naturally produced by
inflation.  Such a model is probably too contrived to be plausible.



\vspace{3mm}
\begin{center}
\bf Acknowledgments
\end{center}
We would like thank N. Sugiyama for providing us with the
CMB anisotropy power spectra and for helpful discussions
and comments.
We are grateful to M. Sasaki and T. Tanaka for discussions and comments.
One of us (K.Y.) would like to thank Professor J. Silk
and the people at the Center for Particle Astrophysics,
University of California, Berkeley, where many parts of this
work were done for their hospitality.
He would like to thank Professor H. Sato for continuous
encouragement.
This work was supported in part by Ministry of Education
Grant-in-Aid for Scientific Research No. 2841.

\vspace{10mm}
\begin{center}
\bf References
\end{center}

\noindent
Bardeen, J. M.,  Bond, J. R., Kaiser, N., \and Szalay, A. S.
 1986, ApJ, {\bf 304}, 15.
\\
Bennett, C.L, et al. 1994, ApJ, {\bf 436}, 423.
\\
Bucher, M., Goldhaber, A. S., \and Turok, N. 1994, PUTP-1507,
  hep-ph/9411206.
\\
Bucher, M., \and Turok, N. 1995, PUTP-1518, hep-ph/9503393.
\\
Bunn, E.F. 1995a, Ph.D. dissertation, U.C. Berkeley Physics Department.
\\
Bunn, E.F. 1995b, in preparation.
\\
Bunn, E.F., Scott, D., \& White, M. 1995, ApJ, {\bf 441}, L9.
\\
Bunn, E.F., \& Sugiyama, N. 1995, ApJ, in press.
\\
Coleman, S., \and De Luccia, F. 1980, Phys. Rev. {\bf D21}, 3305.
\\
Dekel, A. 1994 ARA $\&$ A, 32, 371.
\\
G\'orski, K. M. 1994, ApJ, {\bf 430}, L85.
\\
G\'orski, K. M., Ratra, B., Sugiyama, N., \and Banday, A. J.
  1995, astro-ph/9502034, ApJ, in press.
\\
Gottl\"ober, S., M\"uchet, J. P., \and Starobinsky, A. A. 1994,
  ApJ, {\bf 434}, 417.
\\
Hamazaki, T., Sasaki, M., Tanaka, T., \and Yamamoto, K. 1995,
  KUNS1340.
\\
Kamionkowski, M., Ratra, B., Spergel, D. N., \and Sugiyama, N.
  1994, ApJ, {\bf 434}, L1.
\\
Kashlinsky, A., Tkachev, A. A., \and Frieman, J. 1994,
  Phys. Rev. Lett., {\bf 73}, 1582.
\\
Kodama, H., \and Sasaki, M. 1984, Prog. Theor. Phys. Suppl. {\bf 78}, 1.
\\
Liddle, A. R., Lyth, D. H., Roberts, D., \and Voana, P. T. 1995,
  SUSSEX-AST 95/6-2, astro-ph/9506091, MNRAS, in press.
\\
Linde, A. 1995, preprint SU-ITP-95-5, hep-th/9503097,
  Phys. Lett. {\bf B}, in press.
\\
Linde, A., \and Mezhlumian, A. 1995, preprint SU-ITP-95-11,
  astro-ph/9506017.
\\
Lyth, D. H., \and Stewart, E. D. 1990,  Phys. Lett. {\bf B252}, 336.
\\
Lyth, D. H., \and Woszczyna, A. 1995, Lancaster preprint, astro-ph/9408069.
\\
Mukhanov, V. F., Feldman, H. A., \and Brandenberger, R. H. 1992,
  Phys. Rep. {\bf 215}, 203.
\\
Ostriker, J. P.,  \and Steinhardt, P. J. 1995, astro-ph/9505066.
\\
Peacock, J. A., \and Dodds, S. J. 1994, MNRAS, 267, 1020.
\\
Peebles, P. J. E. 1993, {\it Principle of Physical Cosmology},
  Princeton Univ. Press.
\\
Peter, P., Polarski, D., \and Starobinsky, A. A. 1994, Phys. Rev.
  {\bf D50}, 4827.
\\
Ratra, B., et al. 1995, in preparation.
\\
Ratra, B., \and Peebles, P. J. E. 1994a, ApJ, {\bf 432}, L5.
\\
Ratra, B., \and Peebles, P. J. E. 1994b, preprint PUPT-1444.
\\
Sasaki, M., Tanaka, T., Yamamoto, K., \and Yokoyama, J. 1994,
  Prog. Theor. Phys. 90, 1019.
\\
Sasaki, M., Tanaka, T., \and Yamamoto, K. 1995, Phys. Rev.
  {\bf D51}, 2979.
\\
Scott, D., Silk, J., \and White, M. 1995, preprint.
\\
Smoot, G. F. et al. 1995, ApJ. {\bf 396}, L1.
\\
Stompor, R., G\'orski, K. M., \and Banday, A. J. 1995, preprint.
\\
Sugiyama, N. 1995, preprint CfPA-TH-94-62.
\\
Sugiyama, N., \and Gouda, N. 1992, Prog. Theor. Phys., 88, 803.
\\
Sugiyama, N., \and Silk, J. 1994, Phys. Rev. Lett., {\bf 73}, 509.
\\
Tanaka, T., Sasaki, M., \and Yamamoto, K. 1994,
  Phys. Rev. {\bf D49}, 1039.
\\
Tanaka, T., \and Sasaki, M. 1994, Phys. Rev. {\bf D50}, 6444.
\\
Tegmark, M., \& Bunn, E. F. 1995, Berkeley preprint.
\\
Weinberg, S. 1989, Rev. Mod. Phys., 61, 1.
\\
White, M., \and Bunn, E. F. 1995, CfPA-95-TH-02, ApJ, in press.
\\
White, M., Scott, D., Silk, J., \& Davis, M. 1995, preprint.
\\
Yamamoto, K., Sasaki, M., \and Tanaka, T. 1995, KUNS-1309,
  ApJ, in press.
\\
Yamamoto, K., Tanaka, T., \and Sasaki, M. 1995, Phys. Rev.
  {\bf D51}, 2968.

\def\Q{{Q}}
\def\muK{{\mu {\rm K}}}
\vspace{12mm}
\centerline{Table 1}
\begin{centerline}
{Confidence levels for Bunch-Davies open model }
\end{centerline}
\begin{center}
\begin{tabular}{c@{\hspace{1pc}}c}
\hline\hline
 {\ \ $\Omega_0(>)$ \ \ }&{confidence level $(\%)$}
\\
\hline
$ 0.1$ & $99.4$ \\
$ 0.2$ & $98.4$ \\
$ 0.3$ & $96.2$ \\
$ 0.4$ & $91.8$ \\
$ 0.5$ & $83.2$ \\
\hline
\end{tabular}
\end{center}

\vspace{2mm}
\centerline{Table 2(a)}
\begin{centerline}
{Amplitude of density perturbation for Bunch-Davies open model }
\end{centerline}
\begin{center}
\begin{tabular}{c@{\hspace{1pc}}c@{\hspace{1pc}}c@{\hspace{1pc}}c@{\hspace{1pc}}c}
\hline\hline
 {\ \ $\Omega_0$ \ \ }&{$\cal A$} & {\ \ } & {$\sigma_8$} & {\ \ }
\\
 {\ \ } & {$(h^{-1}{\rm Mpc})^3$} & {h=0.5} & {h=0.65} & { h=0.8}
\\ \hline
$ 0.05$ & $1.96(\Q/27.9\muK)^2~\times10^{2}$  & $0.0041$ & $0.011$& $0.021$ \\
$ 0.1 $ & $2.42(\Q/28.8\muK)^2~\times10^{2}$  & $0.032$ & $0.063$ & $0.099$ \\
$ 0.2 $ & $2.83(\Q/27.8\muK)^2~\times10^{2}$  & $0.15$  & $0.25$  & $0.35$ \\
$ 0.3 $ & $2.94(\Q/25.8\muK)^2~\times10^{2}$  & $0.31$  & $0.48$  & $0.66$ \\
$ 0.4 $ & $2.90(\Q/23.4\muK)^2~\times10^{2}$  & $0.49$  & $0.74$  & $1.00$ \\
$ 0.5 $ & $2.72(\Q/21.1\muK)^2~\times10^{2}$  & $0.69$  & $1.01 $ & $1.33$ \\
$ 0.6 $ & $2.43(\Q/19.3\muK)^2~\times10^{2}$  & $0.89$  & $1.26 $ & $1.63$ \\
$ 0.7 $ & $2.03(\Q/18.3\muK)^2~\times10^{2}$  & $1.05$  & $1.45 $ & $1.90$ \\
$ 0.8 $ & $1.54(\Q/18.3\muK)^2~\times10^{2}$  & $1.20$  & $1.66 $ & $2.10$ \\
$ 0.9 $ & $1.00(\Q/18.9\muK)^2~\times10^{2}$  & $1.31$  & $1.80 $ & $2.25$ \\
\hline
\end{tabular}
\end{center}
\newpage
\centerline{Table 2(b)}
\begin{centerline}
{Amplitude of density perturbation for conformal vacuum open model }
\end{centerline}
\begin{center}
\begin{tabular}{c@{\hspace{1pc}}c@{\hspace{1pc}}c@{\hspace{1pc}}c@{\hspace{1pc}}c}
\hline\hline
{\ \ $\Omega_0$ \ \ } & {$\cal A$} & {\ \ } & {$\sigma_8$} & {\ \ }
\\
 {\ \ } & {$(h^{-1}{\rm Mpc})^3$} & {h=0.5} & {h=0.65} & { h=0.8}
\\ \hline
$ 0.05$ & $2.45(\Q/18.7\muK)^2~\times10^{2}$ & $0.0045$ &$0.012$ & $0.023$ \\
$ 0.1 $ & $2.54(\Q/23.1\muK)^2~\times10^{2}$ & $0.032$  & $0.064$ & $0.10$ \\
$ 0.2 $ & $2.89(\Q/26.5\muK)^2~\times10^{2}$ & $0.15$  & $0.25$  & $0.36$ \\
$ 0.3 $ & $3.01(\Q/25.9\muK)^2~\times10^{2}$ & $0.31$  & $0.49$  & $0.67$ \\
$ 0.4 $ & $2.96(\Q/23.5\muK)^2~\times10^{2}$ & $0.50$  & $0.75$  & $1.01$ \\
$ 0.5 $ & $2.77(\Q/20.6\muK)^2~\times10^{2}$ & $0.69$  & $1.02$  & $1.34$ \\
$ 0.6 $ & $2.46(\Q/18.3\muK)^2~\times10^{2}$ & $0.88$  & $1.27$  & $1.64$ \\
$ 0.7 $ & $2.04(\Q/17.2\muK)^2~\times10^{2}$ & $1.06$  & $1.50$  & $1.90$ \\
$ 0.8 $ & $1.55(\Q/17.5\muK)^2~\times10^{2}$ & $1.20$  & $1.67$  & $2.10$ \\
$ 0.9 $ & $1.00(\Q/18.7\muK)^2~\times10^{2}$ & $1.31$  & $1.80$  & $2.25$ \\
\hline
\end{tabular}
\end{center}

\vspace{5mm}
\centerline{Table 2(c)}
\begin{centerline}
{Amplitude of density perturbation for Harrison-Zel'dovich $\Lambda$-model }
\end{centerline}
\begin{center}
\begin{tabular}{c@{\hspace{1pc}}c@{\hspace{1pc}}c@{\hspace{1pc}}c}
\hline\hline
 {\ \ $\Omega_0$ \ \ } & {\ \ } & {$\sigma_8$} & {\ \ }
\\
 {\ \ } & {h=0.5} & {h=0.65} & { h=0.8}
\\ \hline
$ 0.1 $ & $0.15$ & $0.29$ & $0.46$ \\
$ 0.2 $ & $0.41$ & $0.68$ & $0.98$ \\
$ 0.3 $ & $0.65$ & $1.0$ & $1.4$ \\
$ 0.4 $ & $0.85$ & $1.3$ & $1.7$ \\
$ 0.6 $ & $1.2 $ & $1.7$ & $2.2$ \\
\hline
\end{tabular}
\end{center}

\vspace{5mm}
\centerline{Table 3}
\begin{centerline}
{Large scale bulk velocity for Bunch-Davies open model }
\end{centerline}
\begin{center}
\begin{tabular}{c@{\hspace{1pc}}c@{\hspace{1pc}}c@{\hspace{1pc}}c}
\hline\hline
 {\ \  $\Omega_0$ \ \ } & {\ \ } &
 {$v_{R=40h^{-1}{\rm Mpc}} ({\rm km/s})$ } & {\ \ }
\\
 {\ \ \ \ } & {h=0.5} & {h=0.65} & { h=0.8}
\\ \hline
$ 0.05$ & $6.3$ & $8.6$ & $ 11$ \\
$ 0.1 $ & $ 21$ & $ 29$ & $ 37$ \\
$ 0.2 $ & $ 71$ & $ 90$ & $106$ \\
$ 0.3 $ & $130$ & $160$ & $180$ \\
$ 0.4 $ & $200$ & $230$ & $260$ \\
$ 0.5 $ & $260$ & $300$ & $330$ \\
$ 0.6 $ & $320$ & $360$ & $390$ \\
$ 0.7 $ & $370$ & $410$ & $450$ \\
$ 0.8 $ & $410$ & $450$ & $490$ \\
$ 0.9 $ & $440$ & $480$ & $510$ \\
\hline
\end{tabular}
\end{center}

\newpage
\begin{center}
\underline{Figure Captions}
\end{center}

\noindent
{Figure 1.}
Power spectra of the CMB temperature anisotropy
$l(l+1)C_l\times10^{10}/2\pi$ for (a) the Bunch-Davies vacuum
open model with $\Omega_0=0.1,0.2,0.3,0.4,0.5$.
The data were provided by N. Sugiyama.
These theoretical curves are normalized by the COBE likelihood.
The curves are in descending order of $\Omega_0$ as
one moves down at $l=50$.
The results of several CMB experiments are also shown,
taken from the paper by Scott, Silk, \and White (1995).
To compare with degree-scale observations, careful
investigations are required (Ratra et al. 1995).

\vspace{3mm}
\noindent
{Figure 1(b).} CMB power spectra for the conformal vacuum
open model (G\'orski et al. 1995).  The curves and points are as in
Figure 1(a).

\vspace{3mm}
\noindent
{Figure 1(c).} Power spectrum of the CMB temperature anisotropy
$l(l+1)C_l\times10^{10}/2\pi$ for the Harrison-Zel'dovich
$\Lambda$-model with $\Omega_0=0.1,0.2,0.3,0.4,1.0$.
(Sugiyama 1995)

\vspace{3mm}
\noindent
{Figure 2.} Contour plot of the likelihood function $L(\Omega_0,Q)$
for the Bunch-Davies open model. The contour range is from
$L=0.25$ to $L=1.5$, where the likelihoods are scaled
so that $L=1$ corresponds to a flat Harrison-Zel'dovich
spectrum with maximum-likelihood normalization.

\vspace{3mm}
\noindent
{Figure 3.} The marginal likelihood $L_{\rm marg}(\Omega_0)$
as a function of $\Omega_0$ for both
the Bunch-Davies open model (solid line)
and the conformal vacuum open model (dashed line).

\vspace{3mm}
\noindent
{Figure 4(a).} Power spectrum of density perturbation $a_0^3 P(k)$
for the open model with  $\Omega_0=0.1$, $0.2$, $0.3$, $0.4$, $0.5$, for
$h=0.75, 0.70, 0.65, 0.65, 0.60$, respectively.
We have taken $\Omega_Bh^2=0.0125$.
The points are from Peacock \and Dodds (1994).

\vspace{3mm}
\noindent
{Figure 4(b).} Same figure as Fig.4(a) but for the Harrison-Zel'dovich
$\Lambda$-model with $\Omega_0=$ $0.1$, $0.2$, $0.3$, $0.4$, $1.0$.
We have taken $\Omega_Bh^2=$$0.0128$ and $h=0.8$.

\vspace{3mm}
\noindent
{Figure 5(a).} Contours of $\sigma(M_R=10^{12} {\rm M}_\odot,z=1)$,
in the $(\Omega_0-h)$ plane for the Bunch-Davies open models.
The contour range is from
$\sigma=1.0$ to $\sigma=6.0$. The dashed lines are $\sigma=1.6$ and
$2.4$.

\vspace{3mm}
\noindent
{Figure 5(b).} Same figure as Fig.5(a) but for
Harrison-Zel'dovich $\Lambda$-models.

\end{document}